\documentclass[]{article}
\usepackage{graphicx}
\usepackage{float}
\usepackage{color, colortbl}
\usepackage{xcolor}
\usepackage{array}
\usepackage{multirow}
\usepackage{footnote}
\usepackage{cite}
\begin{document}

\title{Tech Report: CSMA/ECA in Non-Saturation Scenarios}
\author{Luis Sanabria-Russo}
\date{\today}
\maketitle

\begin{abstract}
During the past month we have been trying to understand both the behavior of CSMA/ECA and how our simulator works. In this report I try to asses some of those doubts and provide a groundwork for discussion of past and new ideas for further develop the MAC protocol.
\end{abstract}

\section{What was the problem?}\label{problem}
CSMA/ECA instruct nodes to get out of the contention when their MAC queue empties. These nodes $-$ upon receiving a new packet in the MAC queue $-$ will rejoin the contention picking a random backoff counter $B\in[0,CW_{\min}]$.

This behavior brings again a transitory state, where nodes that just rejoined the contention will attempt to make a successful transmission and get back into the collision-free state.

Our first attempts to take a look at what happens under this condition were unsuccessful mainly for the lack of code considering this scenario, i.e.: packet erasure from the MAC queue upon successful transmissions or retransmission attempts, delay measures and the determination of when was it really an unsaturated system.

Nevertheless, those times are behind us. In the following sections we will take a look at some of the results, namely: 
	\begin{itemize}
		\item throughput,
		\item delay,
		\item number of packets in the MAC queue at the end of each simulation,
		\item number of times a MAC queue empties,
		\item average backoff stage at the end of each simulation, and
		\item average collisions.
	\end{itemize}

\section{Throughput in non-saturation}

\begin{figure}[htbp]
	\centering
	\includegraphics[width=0.7\linewidth, angle=-90]{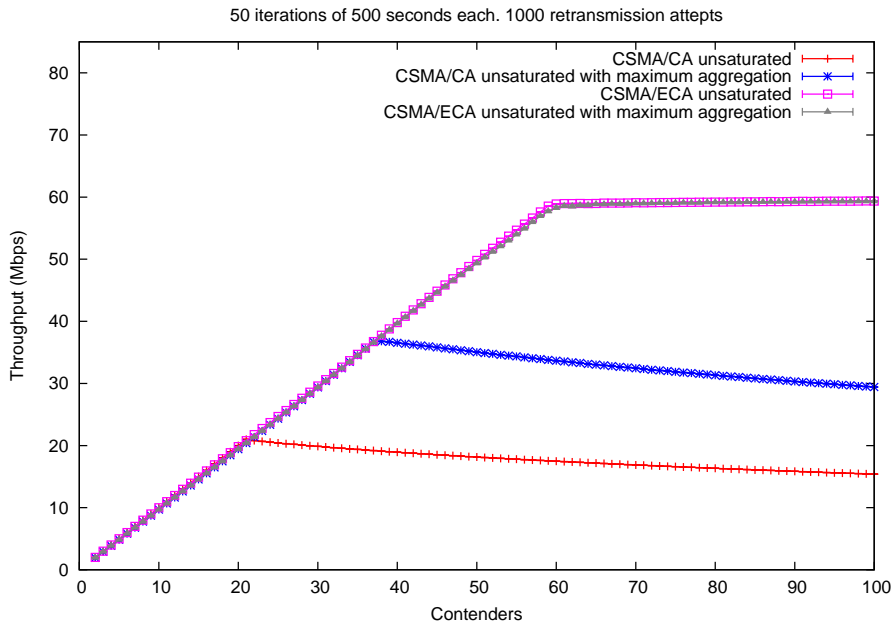}
	\caption{Throughput in non-saturation
	\label{unsat:throughput}}
\end{figure}

We can see form  Figure~\ref{unsat:throughput} that as each tested protocol enters in saturation, it deviates from the straight line (which is the offered traffic load).

\section{Delay}

The delay metric that we are using in this report measures the time elapsed between the moment a packet enters the MAC queue, up until an acknowledgement for this (these) packet(s) is received.

CSMA/ECA offers lower delay than CSMA/CA, nevertheless we can see an otherwise strange behavior at around 60 nodes. What happens?

\begin{figure}[htbp]
	\centering
	\includegraphics[width=0.7\linewidth, angle=-90]{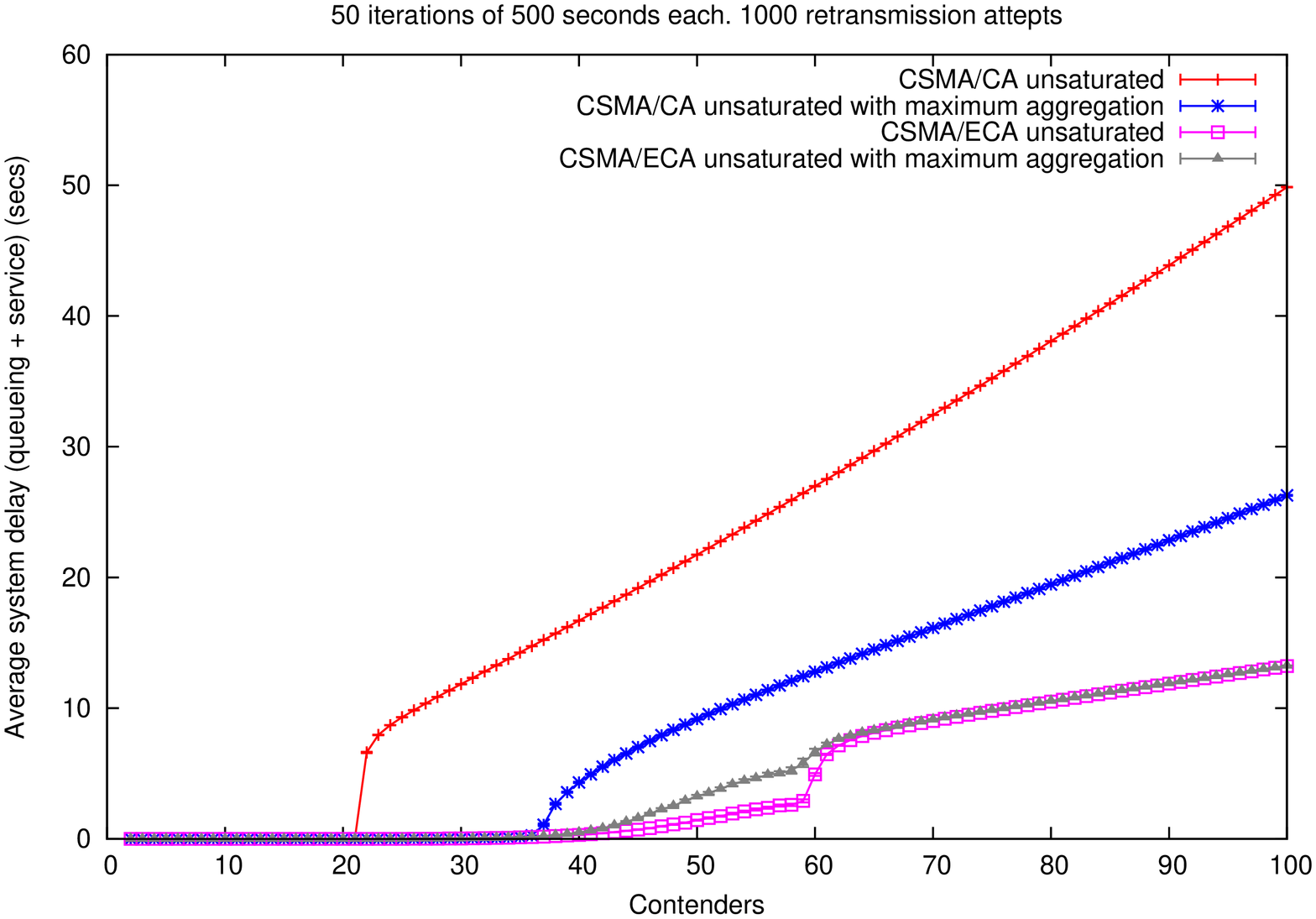}
	\caption{Delay
	\label{unsat:delay}}
\end{figure}

\begin{figure}[htbp]
	\centering
	\includegraphics[width=0.7\linewidth, angle=-90]{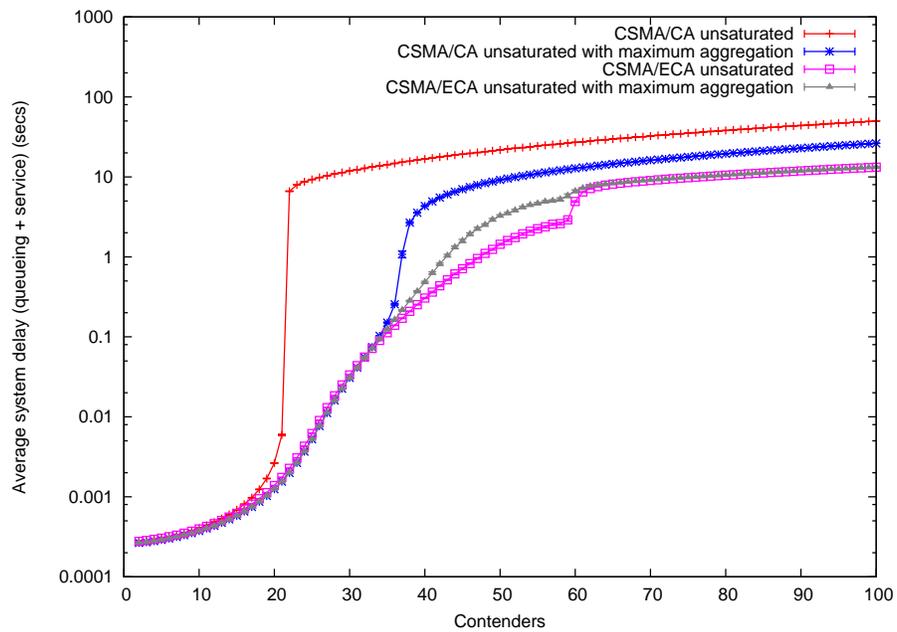}
	\caption{Delay with logscale view
	\label{unsat:delayLog}}
\end{figure}

If we take a closer look at what happens with the delay curves (using a logarithmic scale), the "\emph{bumb}" appears again at 60 nodes.

This bumb is related to the saturation of CSMA/ECA. 

We can see from Figure~\ref{unsat:delayLog}, that this behavior is similar to CSMA/CA when it is about to enter in saturation. If we use the "CSMA/CA unsaturated with maximum aggregation" curve as an example:

	\begin{itemize}
		\item From around 20 to around 35 contenders, this curve seems to be concave; as both CSMA/ECA curves are from around 20 to 60 contenders.
		\item The concavity ends at around 35 contenders (which coincides with the saturation point, see Figure~\ref{unsat:throughput}), just to change to another concavity from the saturation point forward.
		\item For the CSMA/ECA curves this behavior seems to be the same, but this time at around CSMA/ECA's saturation point (around 60 nodes).
	\end{itemize}

\begin{figure}[htbp]
		\centering
		\includegraphics[width=0.7\linewidth, angle=-90]{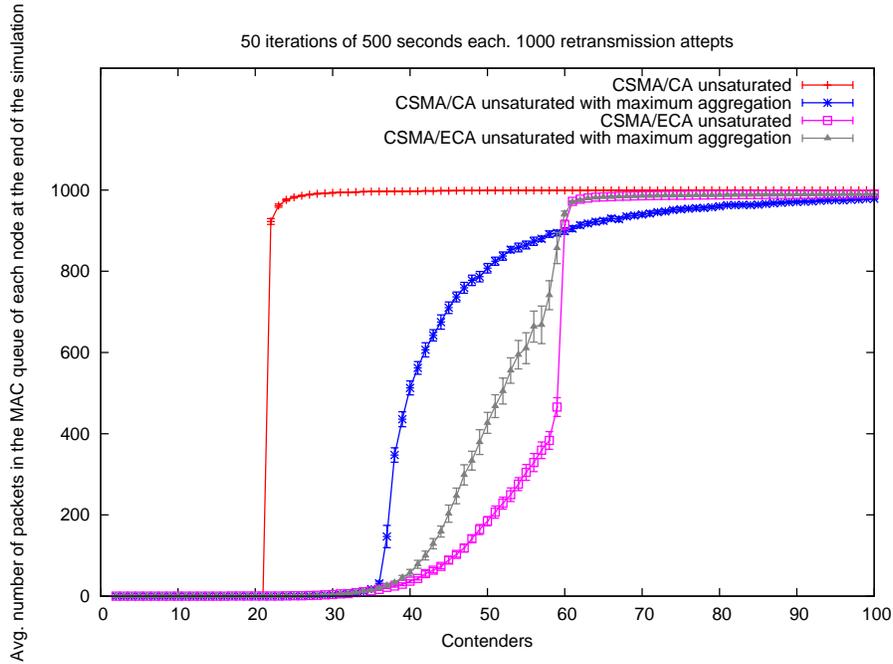}
		\caption{Average MAC queue size per node at the end of each simulation
		\label{unsat:QSize}}
	\end{figure}

This increase in the delay at saturation point will be appreciated as a rapid increase in the number of packets in the MAC queue, see Figure~\ref{unsat:QSize}.

\section{Collisions}
CSMA/ECA eliminates collisions while in saturation and for a limited number of contenders in 802.11 networks. 

Nevertheless, as mentioned in Section~\ref{problem}, collisions may reappear due to the re-joining of nodes after their MAC queues were emptied.

\begin{figure}[htbp]
		\centering
		\includegraphics[width=0.7\linewidth, angle=-90]{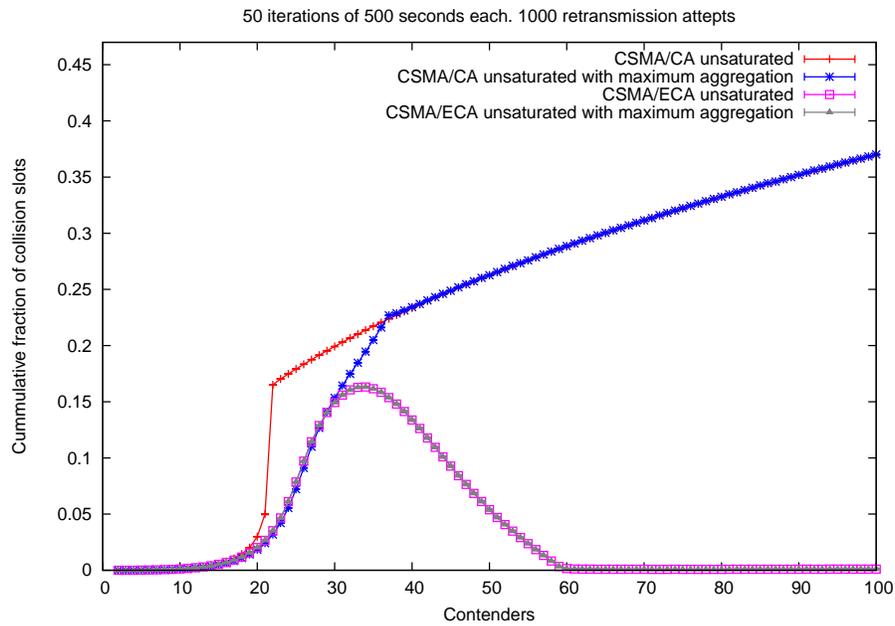}
		\caption{Cummulative fraction of collision slots
		\label{unsat:collisions}}
	\end{figure}

\begin{figure}[htbp]
		\centering
		\includegraphics[width=0.7\linewidth, angle=-90]{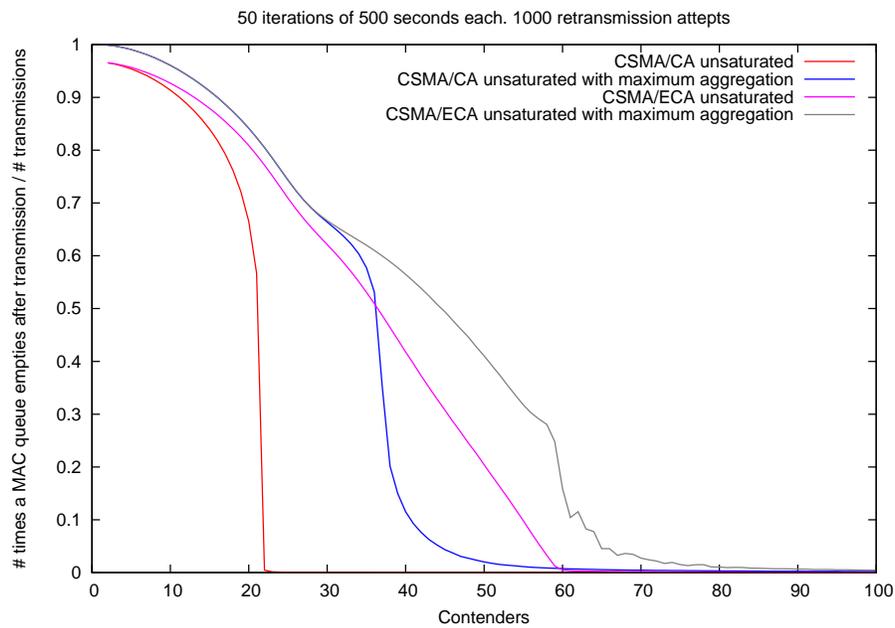}
		\caption{\# Times a MAC queue emties after transmission / \# Transmissions
		\label{unsat:QEmpties}}
	\end{figure}

In Figure~\ref{unsat:collisions} an increase on the average number of collisions is appreciated in the CSMA/ECA curves. This is the result of many nodes emptying their MAC queue and rejoining the contention. Figure~\ref{unsat:QEmpties} provides evidence of this event, where from 20 to 60 nodes CSMA/ECA's nodes empty their MAC queue sometime in the simulation and collide.

\begin{figure}[htbp]
		\centering
		\includegraphics[width=0.7\linewidth, angle=-90]{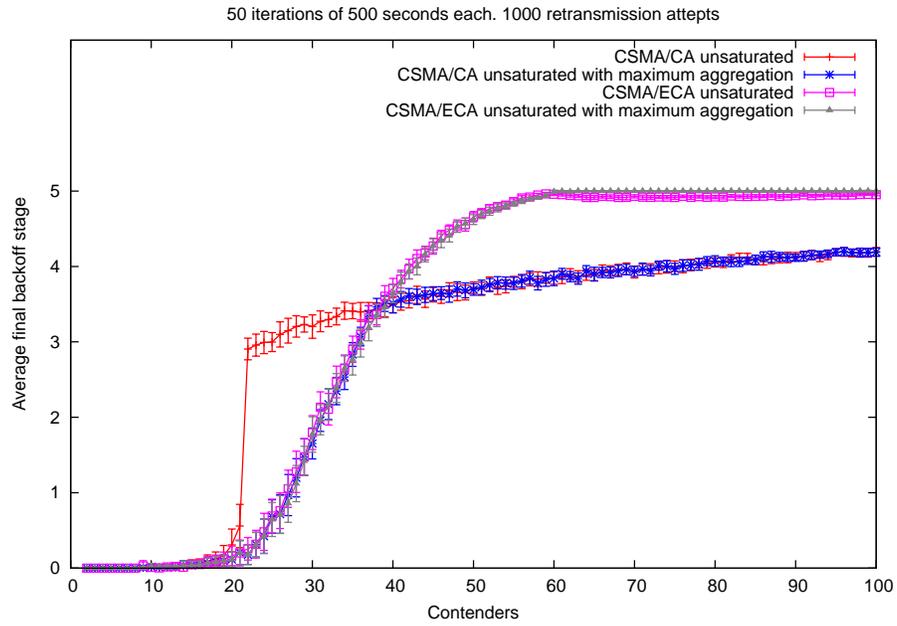}
		\caption{Average backoff stage at the end of each simulation
		\label{unsat:avgBackoffStage}}
	\end{figure}

This also produces an increase in the average backoff stage, which is appreciated in Figure~\ref{unsat:avgBackoffStage}.

\end{document}